\documentclass{article}

\usepackage[utf8]{inputenc}
\usepackage[T1]{fontenc}
\usepackage{arxiv}
\usepackage{amsmath,amsfonts,amssymb}
\usepackage{graphicx}
\usepackage{float}
\usepackage{booktabs}
\usepackage[numbers]{natbib}
\usepackage[colorlinks=true, allcolors=blue]{hyperref}
\usepackage{authblk}
\usepackage{siunitx}
\usepackage{tikz}
\usetikzlibrary{arrows.meta}

\title{Reconstruction-Independent Resolution Limits in Preclinical Cone-Beam Micro-CT: A Closed-Form Analysis}
\renewcommand{\shorttitle}{Resolution Limits in Preclinical Micro-CBCT}

\author[a]{Falk L. Wiegmann}
\author[a,b]{Nancy L. Ford\thanks{Corresponding author: nlford@dentistry.ubc.ca}}
\affil[a]{Department of Physics and Astronomy, The University of British Columbia, Vancouver BC, Canada}
\affil[b]{Department of Oral Biological and Medical Sciences, The University of British Columbia, Vancouver BC, Canada}

\begin{document}

\maketitle

\begin{abstract}
Spatial resolution in preclinical cone-beam micro-CT is bounded by two reconstruction-independent limits, a photon-noise limit from the projection-domain signal-known-exactly, background-known-exactly (SKE/BKE) ideal observer with the Rose criterion, and an angular-sampling limit from the Crowther criterion. The photon-noise bound is a property of the input data and applies to any reconstruction algorithm. The angular-sampling bound applies to classical analytic and iterative reconstructions (FDK, FBP, SIRT) that do not inject signal priors. We derive both limits in closed form for a circular feature in a uniform background, including an off-axis fan-beam extension, and combine them into a spatially varying effective resolution map. The combined limit yields a closed-form scan-design optimum. The maximum worst-case spatial frequency scales as the cube root of the total photon budget per detector pixel (per-view flux times number of views), and an analytic rule sets how to split that budget between flux and views. We illustrate the framework on a representative preclinical micro-CT scanner, with parametric sweeps over flux and projection count that show how each acquisition parameter reshapes the resolution map. An accompanying open-source web calculator implements these bounds for arbitrary scanner geometries.
\end{abstract}

\keywords{micro-CT \and cone-beam CT \and ideal observer \and Rose criterion \and Crowther criterion \and spatial resolution \and detectability}

\section{Introduction}\label{sec:introduction}

The spatial resolution of a CT scanner is usually characterised by image-domain metrics such as the modulation transfer function (MTF), measured from a specific reconstruction of a specific phantom. These metrics are useful, but they conflate two different sources of resolution loss, ones that originate in the data (photon noise, angular undersampling) and ones that originate in the reconstruction (apodization, regularisation, voxel grid). Algorithm-comparison studies, including our own~\cite{wiegmann_fdk_benchmark_2026, wiegmann_filter_sweep_2026}, report MTF values that vary by factors of two across reconstruction settings. None of them answer a more basic question. How much room is left to improve, given the data we already have?

The first bound on the answer is the projection-domain ideal observer, which evaluates the matched-filter detectability index $d'$ of a known feature in known background~\cite{barrett_myers_2004, hsieh_minimum_snr_2022, hanson_detectability_1979}. Under signal-known-exactly and background-known-exactly (SKE/BKE) assumptions, $d'$ is a deterministic function of photon statistics and scan geometry~\cite{chesler_1977, gore_tofts_1978}, with no dependence on the reconstruction algorithm. The Rose criterion ($d' \geq 3$, or 5 in some recent work such as Hsieh et al.~\cite{hsieh_minimum_snr_2022}) converts $d'$ into a minimum detectable feature diameter $d_{\min}$~\cite{faulkner_moores_1984}, and via $f = 1/(2 d_{\min})$ into a spatial-frequency limit. The bound is set by the input data. No reconstruction algorithm of any kind can exceed it.

The second bound is the Crowther criterion~\cite{crowther_1970, kak_slaney_1988, natterer_2001}, which states that to reconstruct an image of radius $r$ with spatial-frequency content up to $f_{\max}$ requires at least $N_\theta = 2\pi r f_{\max}$ angular samples. For a fixed number of projections $N_\theta$ acquired over an arc $\Delta\theta_{\text{span}}$, the spatial-frequency limit at radial distance $r$ from the rotation axis is $f_{\max}(r) = N_\theta / (2 \Delta\theta_{\text{span}} \, r)$. This is a Nyquist bound on what an image-domain observer can extract from the angular samples without additional prior information about the signal. Classical analytic reconstructions such as filtered backprojection (FBP) and the Feldkamp-Davis-Kress (FDK) cone-beam algorithm, and iterative reconstructions without strong signal priors such as the simultaneous iterative reconstruction technique (SIRT) and ordered-subset simultaneous algebraic reconstruction technique (OS-SART), inherit it. Reconstructions that incorporate priors, such as neural networks trained on relevant data, implicit neural representations, and deep image priors~\cite{zang_intratomo_2021, ruckert_neat_2022, shen_nerp_2022, ulyanov_dip_2018, mildenhall_nerf_2020}, are not bound by Crowther and have been shown to recover frequencies above it. The cost is that they invoke assumptions about the signal that the Crowther bound does not.

Both bounds sit inside the broader cascaded-systems framework of medical-physics image quality~\cite{rabbani_cascade_1987, cunningham_cascaded_1994, tward_cascaded_2008}, which propagates signal and noise through every stage of the imaging chain to produce a frequency-dependent noise-equivalent quanta NEQ($f$) and a task-based detectability index $d'^2 = \int |W(f)|^2 \,\mathrm{NEQ}(f)\, df$. The matched-filter SKE/BKE bound used here is the linear shift-invariant (LSI), stationary-Gaussian, no-anatomical-clutter limit of that framework. Specifically, it is the projection-domain limit of the cascaded NEQ in which detector blur, electronic noise, and reconstruction filtering have not yet entered the chain, leaving the input photon count $N_0$ as the only noise-relevant input. The $d^3$ disc-detection scaling we derive analytically below is the same scaling Gang et al.~\cite{gang_anatomical_2010} obtain via cascaded NEQ in cone-beam CT. We chose the closed-form simplification for three reasons. It is evaluable from a handful of measurements rather than a full system MTF and noise-power-spectrum (NPS) characterisation. It exposes scan-design trade-offs in analytic form. And it keeps the angular-sampling constraint (Crowther) separate from the photon-noise constraint, which the integrated NEQ formulation tends to fold together.

This paper combines both limits into a single, spatially varying picture. We derive each bound for a circular feature in a uniform background, evaluate them on a GE eXplore CT~120 preclinical micro-CT scanner, and present the combined-limit map as the effective resolution ceiling for any classical reconstruction of these data. We then sweep the scan parameters that the experimenter controls, exposure and projection count, to show how each parameter reshapes the combined-limit surface. Our specific aims are to:
\begin{enumerate}
    \item Derive a closed-form expression for the photon-noise minimum detectable diameter $d_{\min}(\Delta\mu, \mathbf{r})$ as a function of feature contrast and position, including an off-axis fan-beam extension that handles partial-arc geometries.
    \item Compose a combined-limit map $f_{\text{eff}}(\mathbf{r}) = \min[f_{\text{noise}}(\mathbf{r}), f_{\text{Crowther}}(\mathbf{r})]$, with explicit treatment of its scope as a heuristic for the binding limit of classical reconstructions, and identify crossover contours that mark which limit dominates where.
    \item Use parametric sweeps over $N_0$ and $N_\theta$ to characterise how the combined-limit surface deforms with each scan-design choice, including the implications for trading dose against angular sampling.
    \item Derive a closed-form scan-design optimum under a fixed total-dose budget $\Phi = N_0 N_\theta$, with the dose-resolution law $f_{\min}^{\max} \propto \Phi^{1/3}$ and the optimal-allocation ridge $N_\theta^\star \propto \sqrt{N_0}$ (Section~\ref{sec:res_optimal}).
    \item Apply a region-resolved detectability map to a real reconstructed mouse specimen (Section~\ref{sec:disc_design}), showing how the bound can be sampled at every pixel of an FDK reconstruction to inform ROI selection and feature detectability for downstream analysis.
\end{enumerate}
An open-source interactive web calculator~\cite{wiegmann_ideal_observer_calculator_2026} implements these bounds for arbitrary scanner parameters.

\section{Background}\label{sec:background}

\subsection{Projection-domain ideal observer}\label{sec:bg_ideal_observer}

Consider a circular disc of diameter $d$ and linear-attenuation contrast $\Delta\mu$ embedded in a uniform background of attenuation $\mu_{bg}$ and chord length $L$. The unattenuated photon count per detector pixel is $N_0$, the transmitted count through the background is $N_{bg} = N_0 e^{-\mu_{bg} L}$, and for $N_{bg} \gg 1$ the photon noise variance per pixel in the log-transformed projection is $\sigma_p^2 = 1/N_{bg}$. A ray at offset $t$ from the disc centre passes through a chord of length $2\sqrt{R^2 - t^2}$ (with $R = d/2$), so the disc contribution to the projection is
\begin{equation}\label{eq:projection_signal}
\Delta p(t) = 2\,\Delta\mu \sqrt{R^2 - t^2}, \quad |t| < R,
\end{equation}
and zero otherwise. Because the disc is circular, $\Delta p(t)$ is identical at every projection angle.

The SKE/BKE ideal observer is the matched filter~\cite{barrett_myers_2004, hsieh_minimum_snr_2022}. Its squared detectability index is the projection-noise-weighted sum of squared signal:
\begin{equation}\label{eq:dprime_sum}
d'^2 = \sum_{\theta_j} \sum_{i} \frac{[\Delta p(t_i)]^2}{\sigma_p^2}.
\end{equation}
Converting the pixel sum to an integral with detector pitch $\Delta a_{obj}$ at the object plane and using $\int_{-R}^{R}(R^2 - t^2)\,dt = 4R^3/3 = d^3/6$:
\begin{equation}\label{eq:dprime_closed}
d'^2 = \frac{2\,N_\theta\,N_{bg}\,\Delta\mu^2\,d^3}{3\,\Delta a_{obj}}.
\end{equation}
Setting $d' = d'_{\text{th}}$ (the detection threshold) gives the minimum detectable disc diameter
\begin{equation}\label{eq:dmin}
d_{\min} = \left( \frac{3\,d'^2_{\text{th}}\,\Delta a_{obj}}{2\,N_\theta\,N_{bg}\,\Delta\mu^2} \right)^{1/3},
\end{equation}
and the corresponding spatial-frequency limit is
\begin{equation}\label{eq:f_noise}
f_{\text{noise}} = \frac{1}{2\,d_{\min}}.
\end{equation}
Throughout we use the classical Rose threshold $d'_{\text{th}} = 3$~\cite{rose_vision_1973}. The stricter $d'_{\text{th}} = 5$ used in some recent work such as Hsieh et al.~\cite{hsieh_minimum_snr_2022} multiplies $d_{\min}$ by $(25/9)^{1/3} \approx 1.40$.

\paragraph{Off-axis (fan-beam) extension.}
For a feature at position $\mathbf{r} = (x, y)$ inside a phantom of radius $R_{ph}$ scanned by a point source at distance $D$ from the isocentre, two quantities in Equation~\ref{eq:dprime_closed} become angle-dependent. The chord through the phantom is
\begin{equation}\label{eq:chord}
\text{chord}(\theta, \mathbf{r}) = 2\sqrt{(\mathbf{S}(\theta) \cdot \hat{\mathbf{d}})^2 - (D^2 - R_{ph}^2)},
\end{equation}
where $\mathbf{S}(\theta) = (D\cos\theta, D\sin\theta)$ is the source position and $\hat{\mathbf{d}} = (\mathbf{r} - \mathbf{S})/|\mathbf{r} - \mathbf{S}|$. The local object-plane pixel pitch also scales with source-to-feature distance,
\begin{equation}\label{eq:da_local}
\Delta a_{obj}(\mathbf{r}, \theta) = \Delta a_{obj} \cdot \frac{|\mathbf{r} - \mathbf{S}(\theta)|}{D}.
\end{equation}
Folding both into Equation~\ref{eq:dprime_closed} gives
\begin{equation}\label{eq:dmin_offaxis}
d_{\min}(\mathbf{r}) = \left( \frac{3\,d'^2_{\text{th}}\,\Delta a_{obj}}{2\,\Delta\mu^2\,D\,\sum_{\theta_j} \dfrac{N_0\,e^{-\mu_{bg}\,\text{chord}(\theta_j, \mathbf{r})}}{|\mathbf{r} - \mathbf{S}(\theta_j)|}} \right)^{1/3},
\end{equation}
which reduces to Equation~\ref{eq:dmin} at the isocentre and for full-arc scans at large $D$. For partial-arc acquisitions the map is no longer radially symmetric.

\subsection{Angular sampling (Crowther criterion)}\label{sec:bg_crowther}

The Crowther criterion~\cite{crowther_1970, kak_slaney_1988} bounds the spatial-frequency content recoverable by a classical reconstruction (FBP, FDK, SIRT) from $N_\theta$ projections over an arc $\Delta\theta_{\text{span}}$. At radial distance $r$ from the rotation axis,
\begin{equation}\label{eq:crowther}
f_{\text{Crowther}}(r) = \frac{N_\theta}{2\,\Delta\theta_{\text{span}}\,r}.
\end{equation}
This is a Nyquist statement on the angular spacing of view samples. Features at radius $r$ are seen by sequential views with an angular offset that, projected to the feature, must be smaller than half the period of the highest recovered frequency. The bound diverges at the isocentre ($r \to 0$) but is capped by the detector Nyquist frequency $f_{\text{det}} = 1/(2\Delta a_{obj})$.

Equation~\ref{eq:crowther} is a bound on what an image-domain observer can extract from the projection-angle samples alone, without prior information about the signal (Section~\ref{sec:introduction} discusses reconstructions that incorporate priors and can therefore exceed it). The photon-noise bound of Section~\ref{sec:bg_ideal_observer} has no such restriction. It operates on the projection data before any reconstruction.

\subsection{Combined effective limit}\label{sec:bg_combined}

For a classical reconstruction, the effective spatial-frequency limit at position $\mathbf{r}$ is the smaller of the two bounds:
\begin{equation}\label{eq:f_eff}
f_{\text{eff}}(\mathbf{r}) = \min\!\left[ f_{\text{noise}}(\mathbf{r}), \; f_{\text{Crowther}}(\mathbf{r}) \right].
\end{equation}
The locus where $f_{\text{noise}} = f_{\text{Crowther}}$ partitions the field of view into a noise-dominated region (typically near the centre of attenuating phantoms, where $N_{bg}$ is small) and an angular-dominated region (typically at the periphery, where $r$ is large).

\section{Methods}\label{sec:methods}

\subsection{Scanner and Acquisition}\label{sec:methods_scanner}

All measurements were obtained on the GE eXplore CT~120 micro-CT scanner at \SI{80}{kVp}, \SI{40}{mA}, with \SI{16}{ms} exposure per frame and no frame averaging. The flat-panel detector comprises $3500 \times 2300$ pixels at \SI{0.0284}{mm} pitch. A short-scan acquisition (\ang{192}, 220 projections) was used throughout. Table~\ref{tab:geometry} lists the geometric parameters.

\begin{table}[H]
\centering
\caption{Scanner geometry parameters for the eXplore CT~120 used throughout.}
\label{tab:geometry}
\small
\begin{tabular}{lc}
\toprule
Parameter & Value \\
\midrule
Detector size (pixels) & $3500 \times 2300$ \\
Detector pixel pitch $\Delta a_{det}$ (mm) & 0.0284 \\
Source-to-detector distance SDD (mm) & 451.5 \\
Source-to-isocentre distance SOD (mm) & 396.4 \\
Magnification $M = $~SDD/SOD & 1.139 \\
Object-plane pixel pitch $\Delta a_{obj}$ (mm) & 0.0249 \\
Scan arc $\Delta\theta_{\text{span}}$ & \ang{192} \\
Number of projections $N_\theta$ & 220 \\
\bottomrule
\end{tabular}
\end{table}

The closed-form bounds in Sections~\ref{sec:res_ideal}--\ref{sec:res_optimal} are evaluated on a uniform water-equivalent cylinder of radius $R = \SI{13}{mm}$, a theoretical stand-in for a typical preclinical mouse specimen that sets the chord length and background attenuation in Equations~\ref{eq:dmin}--\ref{eq:dmin_offaxis}. The only new acquisitions are flat-field projections used to measure $N_0$ (Section~\ref{sec:methods_n0}). The \textit{in vivo} mouse data for the detectability map of Section~\ref{sec:disc_design} are re-used from previously published studies~\cite{ford_respiratory_2025, ford_spie_2023} and reconstructed with the Feldkamp-Davis-Kress (FDK) algorithm.

\subsection{Photon-flux measurement}\label{sec:methods_n0}

The unattenuated photon count per pixel $N_0$ was estimated from two flat-field projections at the operating point of Table~\ref{tab:geometry}, with no object in the beam. After dark/bright correction, air pixels were pooled from across both frames. The variance of the half-difference $(T_1 - T_2)/\sqrt{2}$ removes correlated fixed-pattern noise, leaving photon shot noise. Under Poisson statistics this gives
\begin{equation}\label{eq:n0_estimator}
N_0 = \frac{1}{\text{Var}\!\left[(T_1 - T_2)/\sqrt{2}\right]}.
\end{equation}
We obtained $\text{Var}(T_{\text{air}}) = 2.38 \times 10^{-4}$, giving $N_0 = \num{4203}$~photons/pixel. This is the spatial average across the detector. Variation with heel effect, beam shape, or ray obliquity is absorbed into the average rather than resolved spatially. A pixel-wise flat-field calibration would yield $N_0(\theta_j, \mathbf{r})$ but is not implemented here.

\subsection{Evaluation of the resolution-limit surfaces}\label{sec:methods_eval}

The on-axis and Crowther bounds (Equations~\ref{eq:dmin}, \ref{eq:crowther}) are evaluated directly. The off-axis map (Equation~\ref{eq:dmin_offaxis}) is evaluated by summing $N_{bg}(\theta_j, \mathbf{r}) / |\mathbf{r} - \mathbf{S}(\theta_j)|$ over the 220 projection angles, on a $500\times 500$ grid spanning $\pm\SI{47}{mm}$ around the isocentre. The combined-limit map (Equation~\ref{eq:f_eff}) is the pointwise minimum of the two component maps. For the 3D off-mid-slice rendering in Figure~\ref{fig:3d_extension}b, the in-plane chord is scaled by the cone-beam ray-path factor $L_{3\mathrm{D}}/L_{xy}$. FDK Tuy-condition error is not modelled. The $R = \SI{13}{mm}$ cylinder is placed on the rotation axis throughout, except in Figure~\ref{fig:combined}, where it sits at $(2.71, -5.53)$~mm to match a representative mouse specimen~\cite{ford_respiratory_2025}.

\subsection{Parametric sweeps}\label{sec:methods_sweeps}

Two parametric sweeps characterise how the resolution limit responds to scan-design choices, both on the centred $R = \SI{13}{mm}$ specimen model at $\Delta C = \SI{500}{HU}$:
\begin{itemize}
    \item \textbf{Exposure sweep:} $N_0 \in \{\num{1000}, \num{2000}, \num{4203}, \num{10000}, \num{30000}\}$~photons/pixel at fixed $N_\theta = 220$, spanning roughly an order of magnitude either side of the nominal. Crowther is not included in this sweep, because $N_0$ does not enter the Crowther formula. The figure (Figure~\ref{fig:sweep_n0}) shows only the photon-noise component.
    \item \textbf{Projection-count sweep:} $N_\theta \in \{55, 110, 220, 440\}$ at fixed $N_0 = \num{4203}$~photons/pixel, over the same \ang{192} arc, in factor-of-two steps from sparse-view to well-sampled. Increasing $N_\theta$ widens the Crowther cone (Equation~\ref{eq:crowther}) linearly and improves the photon-noise bound as $N_\theta^{1/3}$ in $d_{\min}$ (Equation~\ref{eq:dmin}). The figure (Figure~\ref{fig:sweep_ntheta}) shows the combined limit so both effects are visible.
\end{itemize}
The \ang{192} short-scan is used throughout to highlight the asymmetry of the off-axis maps. A full \ang{360} acquisition would simply render them symmetric. Each sweep is presented as a two-panel figure, a 3D rendering of the limit surface(s) over the specimen disc and a radial profile of the limit along the $+x$ axis.

\section{Results}\label{sec:results}

To illustrate the framework, we evaluate it on a representative preclinical micro-CBCT scanner, the GE eXplore CT~120 of Section~\ref{sec:methods_scanner}. The specific numerical values in what follows correspond to this acquisition geometry. The closed-form scaling laws (the $d^3$ disc-detection scaling, the $\Phi^{1/3}$ dose-resolution law, and the $\sqrt{N_0}$ optimal-allocation ridge) are general results of the framework itself and apply to any scanner.

\subsection{The photon-noise resolution limit}\label{sec:res_ideal}

We evaluate the closed-form bound (Equation~\ref{eq:dmin}) on a centred uniform water-equivalent cylinder of radius $R = \SI{13}{mm}$, matching the typical scale of a preclinical mouse specimen on this scanner. At the rotation axis every projection ray traverses the full diameter $L = 2R = \SI{26}{mm}$, giving $N_{bg} = N_0 e^{-\mu_{bg} L} = \num{4203} \cdot e^{-0.0219 \cdot 26} \approx \num{2379}$~photons/pixel. Table~\ref{tab:dmin_ideal} lists the minimum detectable disc diameters. At our nominal acquisition the photon-noise spatial-frequency limit at the rotation axis is \SI{0.98}{lp/mm} at $\Delta C = \SI{100}{HU}$ and \SI{2.86}{lp/mm} at \SI{500}{HU}. Both sit well below the hardware ceilings of this system, the detector Nyquist of \SI{20.1}{lp/mm} and the reconstruction-grid Nyquist of \SI{6.67}{lp/mm} for our \SI{75}{\micro\metre} voxels. Photon noise sets the bound at low contrast, not hardware or the reconstruction grid.

\begin{table}[H]
\centering
\caption{Minimum detectable disc diameter and spatial-frequency limit at the rotation axis from the projection-domain ideal observer (Equation~\ref{eq:dmin}, $d'_{\text{th}} = 3$), evaluated on the $R = \SI{13}{mm}$ specimen geometry used throughout this work.}
\label{tab:dmin_ideal}
\small
\begin{tabular}{lcccc}
\toprule
$\Delta C$ (HU) & $\Delta\mu$ (mm$^{-1}$) & $d_{\min}$ (mm) & $f_{\text{noise}}$ (lp/mm) & Feature size ($\mu$m) \\
\midrule
100  & 0.00219 & 0.512 & 0.98 & 512 \\
200  & 0.00438 & 0.322 & 1.55 & 322 \\
500  & 0.01095 & 0.175 & 2.86 & 175 \\
1000 & 0.02190 & 0.110 & 4.54 & 110 \\
\bottomrule
\end{tabular}
\end{table}

\paragraph{Gap to actual reconstructions.}
As a calibration of how far real algorithms sit from this bound, we previously measured $d_{\min}$ for our open-source FDK, ASTRA SIRT~\cite{vanaarle_astra_2016}, and TIGRE OS-SART~\cite{biguri_tigre_2016} on a larger image-quality phantom (water-equivalent path $L = \SI{80}{mm}$, $R = \SI{40}{mm}$)~\cite{wiegmann_filter_sweep_2026, wiegmann_fdk_benchmark_2026}. Recomputing the ideal-observer reference at the same geometry, real reconstructions sit a factor of roughly $2.8$ above the bound, and this factor is approximately independent of contrast, consistent with a structural imaging-chain penalty rather than a contrast-specific loss. Two factors drive this gap. The SKE/BKE matched filter is itself optimistic, requiring exact knowledge of signal location and background that no real observer can satisfy. Gang et al.~\cite{gang_anatomical_2010} quantify the penalty incurred by anatomical clutter in cone-beam CT, with the size depending strongly on the task and the acquisition geometry. The rest reflects noise-side propagation in the reconstruction, which closes only with algorithms that better match the matched filter on the noise side, not more dose.

\subsection{Spatial structure of the photon-noise limit}\label{sec:res_3d}

The closed-form bound of Section~\ref{sec:res_ideal} evaluates the photon-noise limit only at the rotation axis. The off-axis fan-beam extension (Equation~\ref{eq:dmin_offaxis}) generalises it to the full field of view, and the same expression extends naturally to 3D using the cone-beam chord scaling described in Section~\ref{sec:methods_eval}. We apply this on the same $R = \SI{13}{mm}$ specimen geometry. Cone-beam FDK Tuy-condition error is not modelled, as it is sub-percent for the \ang{4} cone half-angle of this scanner.

\begin{figure}[H]
    \begin{center}
    \includegraphics[width=\textwidth]{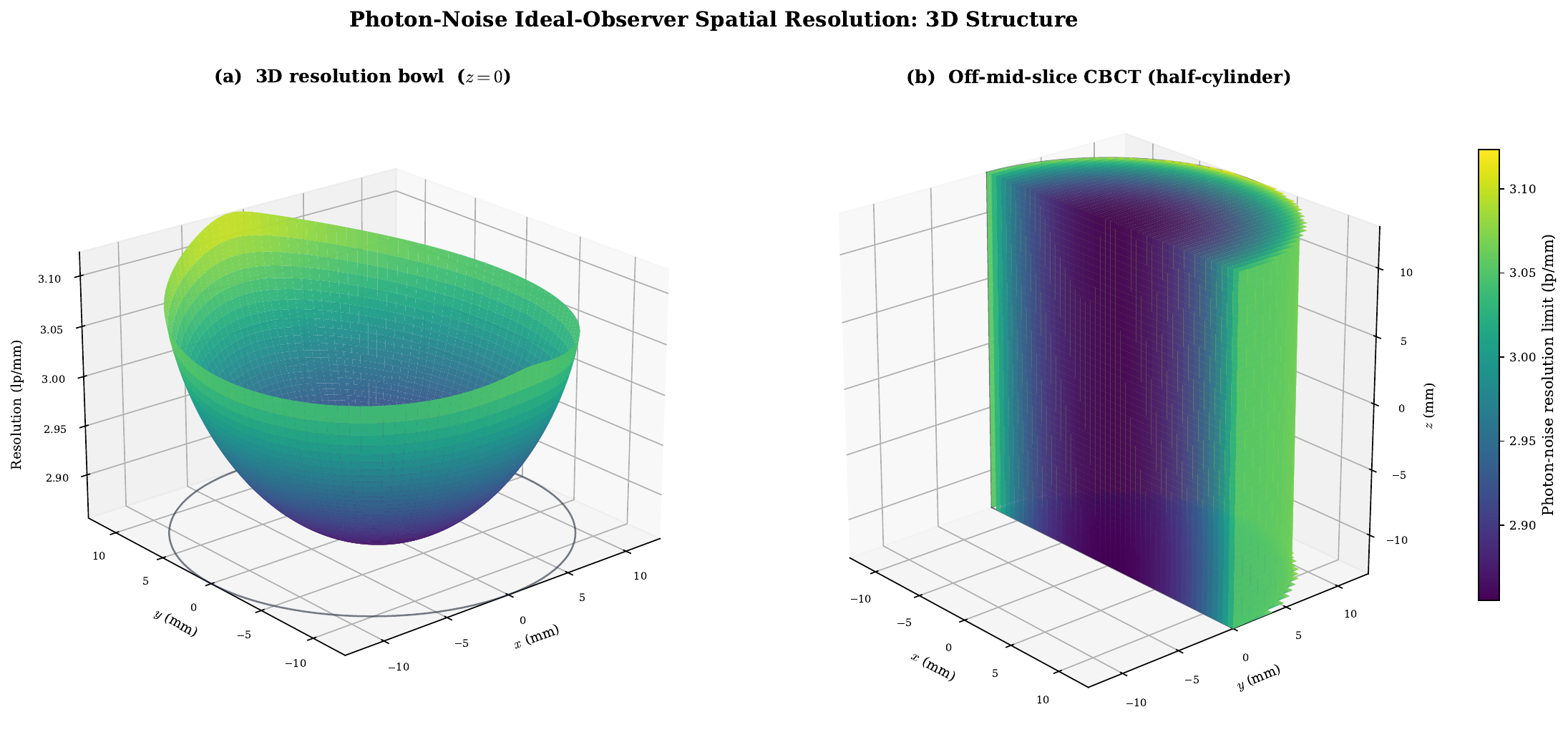}
    \end{center}
    \caption{Three-dimensional structure of the photon-noise ideal-observer resolution limit, at $\Delta C = \SI{500}{HU}$ inside a centred cylindrical specimen model ($R = \SI{13}{mm}$, half-height $Z = \SI{13}{mm}$). The Crowther angular-sampling limit is not included here. The combined limit including it is shown later in Figure~\ref{fig:combined}. (a)~3D resolution bowl on the mid-plane $z = 0$. Surface height encodes $f_{\text{noise}}$ (lp/mm). The specimen rim is outlined at the floor of the plot. (b)~Off-mid-slice half-cylinder rendering: cut plane at $y = 0$ (front face), curved cylinder wall on the $y \geq 0$ half (behind), and top and bottom caps at $z = \pm Z$. For this scanner's \ang{4} cone half-angle the surface is effectively $z$-invariant, so the mid-plane bowl applies throughout the reconstruction volume.}
    \label{fig:3d_extension}
\end{figure}

Figure~\ref{fig:3d_extension}a presents the photon-noise resolution as a 3D surface in the central slice $z = 0$. The familiar ``bowl'' shape is recovered. $f_{\text{noise}}$ is lowest at the specimen centre ($r = 0$), where every projection ray traverses the full diameter and photon attenuation is maximal, and rises toward the rim ($r = R$), where partial-chord rays see less attenuation. For this specimen and contrast, $f_{\text{noise}}$ rises from $\SI{2.86}{lp/mm}$ at the centre of the bowl to $\SI{3.02}{lp/mm}$ at the rim. The centre value matches the at-isocentre closed-form of Table~\ref{tab:dmin_ideal}, confirming that the off-axis formulation reduces to the on-axis bound.

Figure~\ref{fig:3d_extension}b extends the analysis off the mid-plane. The half-cylinder cut renders the resolution on four surfaces of an on-axis cylindrical specimen of half-height $Z = R = \SI{13}{mm}$: the cut plane at $y = 0$, the curved cylinder wall on the $y \geq 0$ side, and the top and bottom caps at $z = \pm Z$. For this scanner the resolution is effectively $z$-invariant across the full volume. At $(r, z) = (0, \pm Z)$ the resolution differs from the central-slice value at the same $r$ by less than \SI{1}{\percent}. The mid-plane analysis therefore applies throughout the reconstruction volume to within sub-percent error. Scanners with larger cone angles, such as wide-detector clinical CBCT systems with cone half-angles of \ang{10}--\ang{15}, would require adding the FDK Tuy-condition penalty omitted here.

\subsection{Effect of exposure on the photon-noise limit}\label{sec:res_sweep_n0}

\begin{figure}[H]
    \begin{center}
    \includegraphics[width=\textwidth]{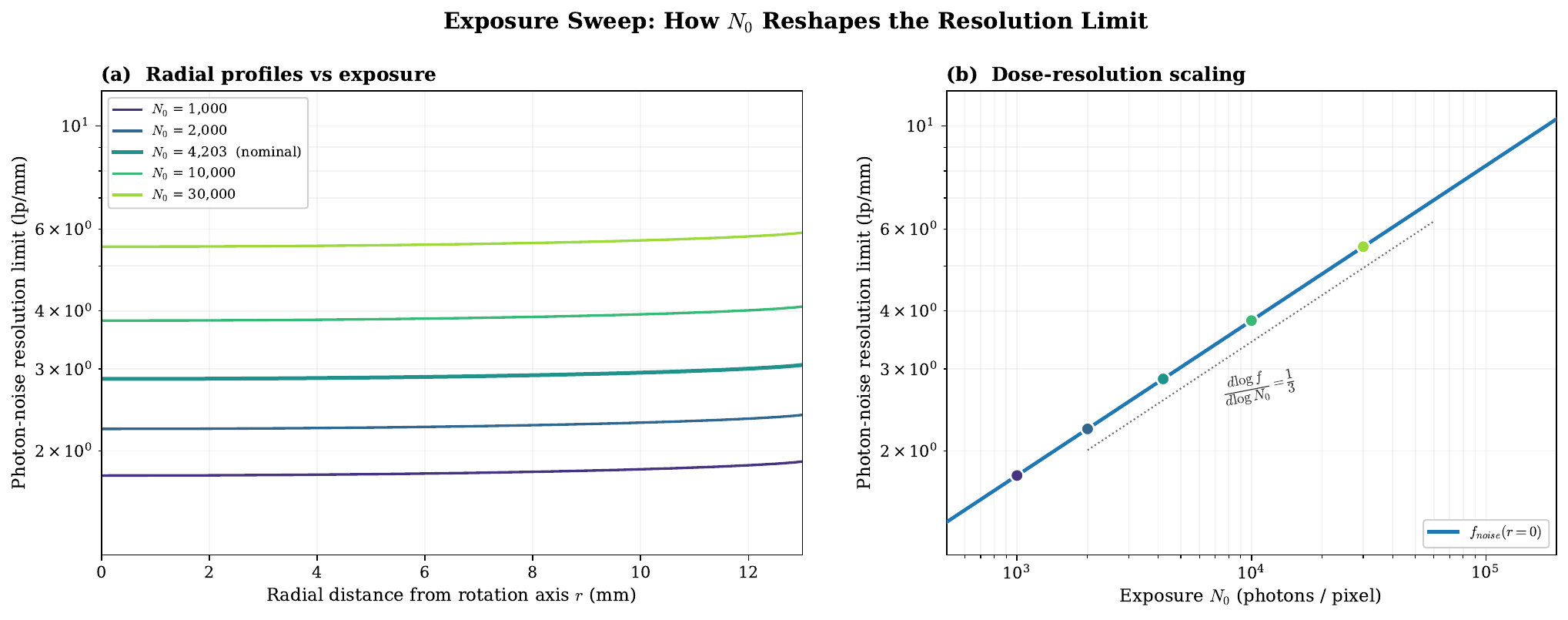}
    \end{center}
    \caption{Effect of exposure $N_0$ on the photon-noise resolution limit at $\Delta C = \SI{500}{HU}$. (a)~Radial profile $f_{\text{noise}}(r)$ at five log-spaced exposures. Increasing $N_0$ raises every curve by the predicted $N_0^{1/3}$ factor without changing its shape. (b)~Log-log dose-resolution scaling at the rotation axis. The blue curve traces $f_{\text{noise}}(r = 0)$. Coloured markers correspond to the five exposures in panel~(a). The dotted guideline shows the predicted $1/3$ slope.}
    \label{fig:sweep_n0}
\end{figure}

Figure~\ref{fig:sweep_n0} characterises how exposure $N_0$ reshapes the photon-noise component of the resolution limit. Figure~\ref{fig:sweep_n0}a overlays the radial profile $f_{\text{noise}}(r)$ for five exposures spanning \num{1000}--\num{30000}~photons/pixel at $\Delta C = \SI{500}{HU}$. Increasing $N_0$ lifts every curve uniformly. Each is a vertical rescaling of the nominal $N_0 = \num{4203}$ profile by the factor $(N_0 / \num{4203})^{1/3}$. The bowl shape is preserved at every exposure. Only its floor moves.

Figure~\ref{fig:sweep_n0}b plots the central-slice scaling on log-log axes. The continuous $f_{\text{noise}}(r = 0)$ curve follows the $N_0^{1/3}$ power law of Equation~\ref{eq:dmin}. A $30\times$ exposure increase delivers a $\sqrt[3]{30} \approx 3.1\times$ resolution gain. The slope guideline confirms the predicted $1/3$ log-log exponent. This is the diminishing-returns regime of the dose-resolution trade. Each octave of dose buys only a $\sqrt[3]{2} \approx 1.26\times$ resolution improvement.

\subsection{From projection domain to reconstruction domain: the angular-sampling ceiling}\label{sec:res_bridge}

The bound discussed so far is the projection-domain photon-noise ceiling. It is the absolute floor that any reconstruction inherits. No algorithm of any kind can resolve a feature whose detectability index falls below the threshold on the projection data themselves. The ceiling is also achievable only by reconstruction algorithms that match the matched filter on the noise side, and the brief comparison in Section~\ref{sec:res_ideal} shows classical FDK and iterative methods operating several times above it.

The same class of reconstructions also faces a second, geometrically motivated ceiling in the image domain, the Crowther angular-sampling limit (Equation~\ref{eq:crowther}). It is a Nyquist statement on the highest spatial frequency recoverable from a finite set of angular samples without prior information about the signal. Analytic methods (FDK, FBP) and iterative methods with weak signal priors (SIRT, OS-SART, OS-EM) inherit it. Reconstructions with strong signal priors do not, as discussed in Section~\ref{sec:introduction} and revisited in Section~\ref{sec:disc_scope}.

The remainder of this section uses the effective resolution limit
\begin{equation*}
f_{\text{eff}}(\mathbf{r}) \;=\; \min\!\left[\,f_{\text{noise}}(\mathbf{r}),\,f_{\text{Crowther}}(\mathbf{r})\,\right]
\end{equation*}
(Equation~\ref{eq:f_eff}) as the ceiling for classical, prior-free reconstructions. Section~\ref{sec:res_crowther} introduces the Crowther map alone, Section~\ref{sec:res_combined} combines it with the photon-noise bound, and Sections~\ref{sec:res_sweep_ntheta}--\ref{sec:res_optimal} explore how scan-design choices reshape the combined surface.

\paragraph{On the formal status of the combined limit.}
The two components of $f_{\text{eff}}$ are not the same kind of object. $f_{\text{Crowther}}$ is a Fourier-domain sampling cap. Above this frequency, no Fourier component can be recovered from $N_\theta$ angular samples without aliasing. It is signal-independent and purely geometric. $f_{\text{noise}} = 1/(2\,d_{\min}(\Delta\mu))$ is something different, a detection threshold for a specific feature (a disc of contrast $\Delta\mu$, at the Rose $d' = 3$ threshold), signal-dependent and SNR-conditioned rather than a frequency cutoff in the strict Fourier sense. The two share units of lp/mm but mean different things. Converting ``smallest detectable feature size'' to ``a frequency'' via $f = 1/(2\,d_{\min})$ is itself a convention. Two independent necessary conditions bind: $f \leq f_{\text{noise}}$ from photon statistics (Section~\ref{sec:bg_ideal_observer}) and $f \leq f_{\text{Crowther}}$ from angular sampling (Section~\ref{sec:bg_crowther}). Their pointwise minimum is the more restrictive of the two at each location. The combined map is a heuristic summary, not a unified theoretical bound. Each component remains rigorous within its own scope.

\subsection{Spatially varying Crowther criterion}\label{sec:res_crowther}

\begin{figure}[H]
    \begin{center}
    \includegraphics[width=\textwidth]{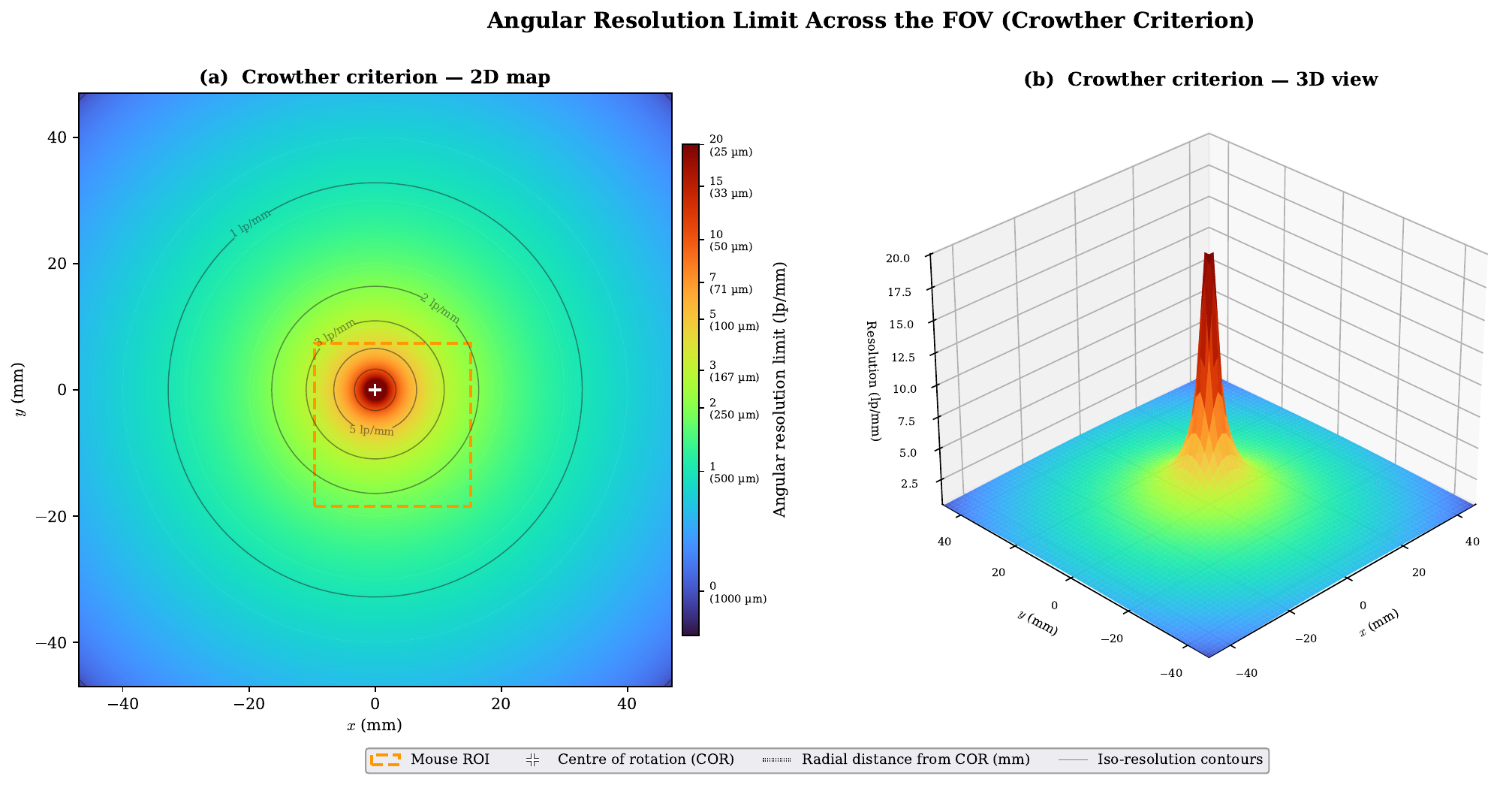}
    \end{center}
    \caption{Angular sampling resolution limit (Crowther criterion). (a)~2D heatmap over the FOV, with iso-resolution contours overlaid in black. (b)~Matching 3D surface. The limit is capped at the detector Nyquist (\SI{20.1}{lp/mm}). Dashed-orange box: representative off-axis mouse ROI. White plus: centre of rotation.}
    \label{fig:crowther}
\end{figure}

Figure~\ref{fig:crowther} renders the Crowther criterion alone as a 2D heatmap and matching 3D surface over the FOV. The limit diverges at the rotation axis and falls as $1/r$. The dashed-yellow ROI marks the location of a representative mouse specimen, illustrating that off-axis specimens see substantially worse angular resolution than ones centred on the isocentre. Across the scanner's circular FOV (radius $\approx \SI{43.6}{mm}$) the Crowther limit ranges from approximately \SI{0.75}{lp/mm} at the rim to the detector-Nyquist cap of \SI{20.1}{lp/mm}. The interactive web calculator~\cite{wiegmann_ideal_observer_calculator_2026} renders this map for arbitrary $N_\theta$, arc coverage, and source distance.

\subsection{Combined effective resolution limit}\label{sec:res_combined}

\begin{figure}[H]
    \begin{center}
    \includegraphics[width=\textwidth]{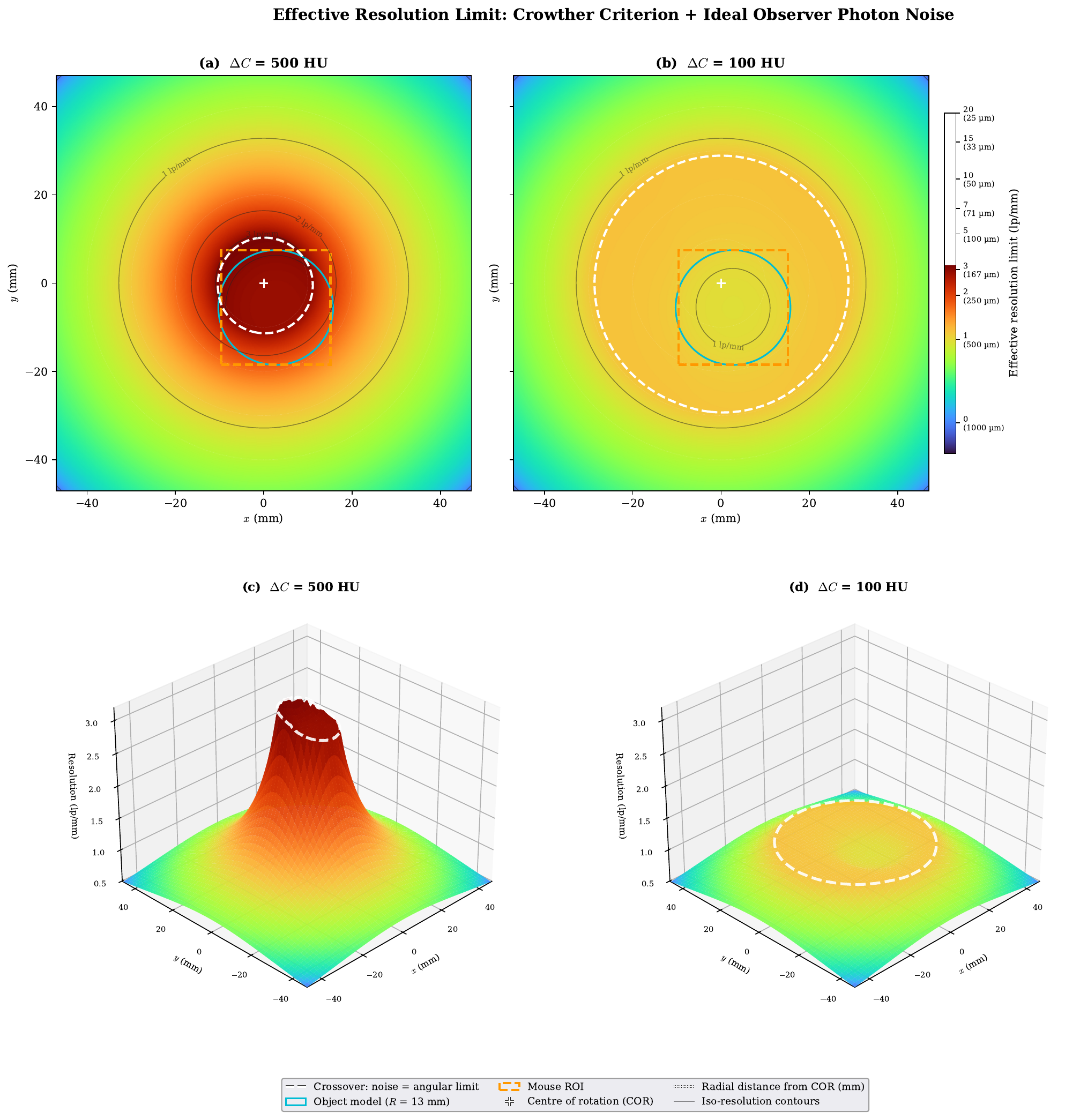}
    \end{center}
    \caption{Combined effective resolution limit $f_{\text{eff}}(\mathbf{r}) = \min[f_{\text{noise}}, f_{\text{Crowther}}]$. (a, c)~$\Delta C = \SI{500}{HU}$. (b, d)~$\Delta C = \SI{100}{HU}$. Top row: 2D heatmaps with iso-resolution contours (black) and noise-Crowther crossover (dashed white). Bottom row: 3D surfaces. Cyan circle: specimen model used in the off-axis ideal-observer evaluation. Dashed-orange box: representative off-axis mouse ROI.}
    \label{fig:combined}
\end{figure}

Figure~\ref{fig:combined} combines the off-axis photon-noise map with the Crowther limit as $f_{\text{eff}} = \min(f_{\text{noise}}, f_{\text{Crowther}})$ (Equation~\ref{eq:f_eff}). The dashed-white contour marks the noise-Crowther crossover. At \SI{500}{HU} the noise-dominated region is a roughly object-shaped bowl centred on the specimen. Outside it, the Crowther limit takes over and the effective resolution decreases radially. At \SI{100}{HU} photon noise dominates almost the entire FOV. The bowl floor has dropped below the Crowther walls everywhere except very near the rotation axis. The 3D surfaces in panels~(c) and~(d) make the ``bowl'' shape explicit, a depression in the centre with rising walls outside it.

\subsection{Parametric sweep: projection count}\label{sec:res_sweep_ntheta}

\begin{figure}[H]
    \begin{center}
    \includegraphics[width=\textwidth]{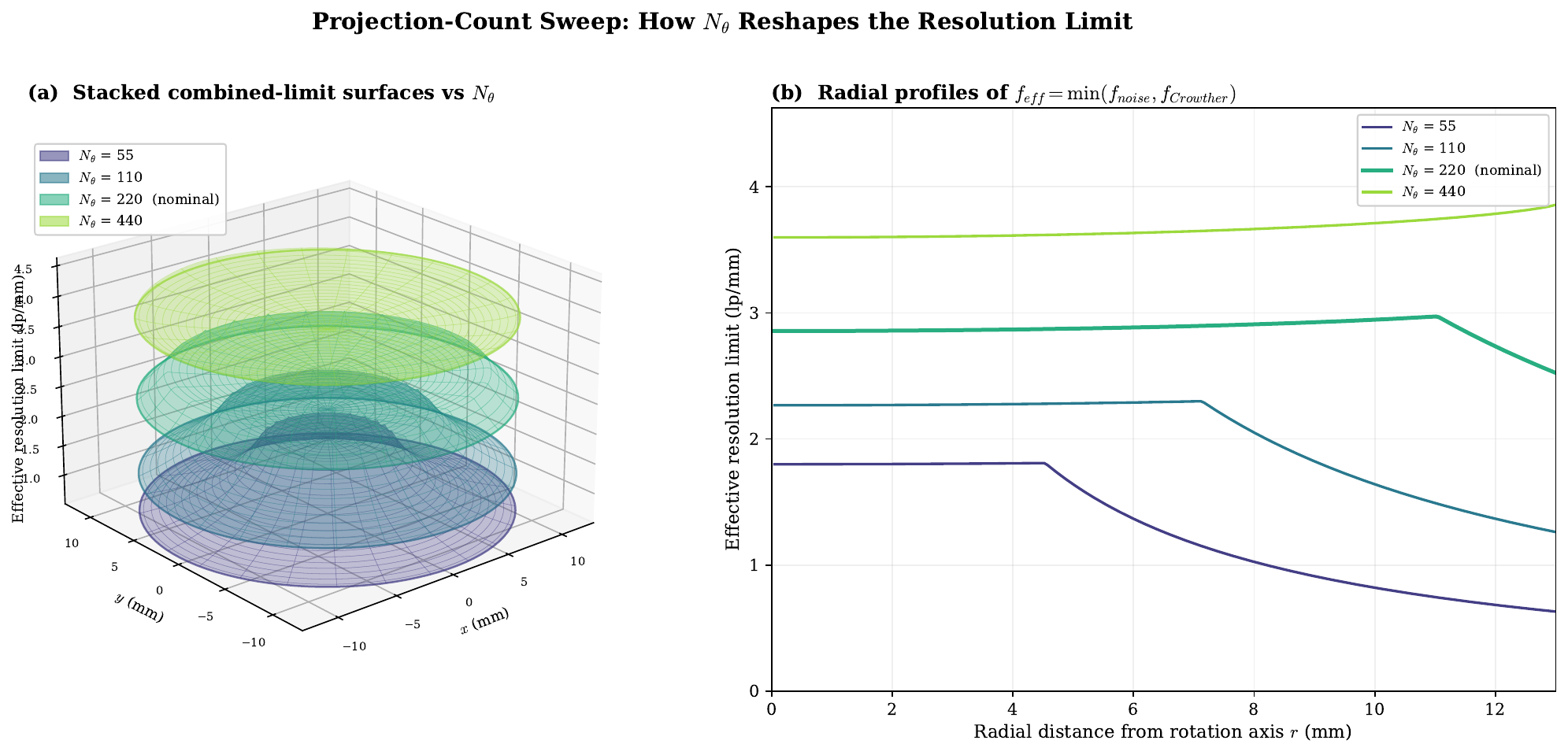}
    \end{center}
    \caption{Effect of projection count $N_\theta$ on the combined-limit resolution surface at $\Delta C = \SI{500}{HU}$. (a)~Stacked 3D surfaces of $f_{\text{eff}}(x, y) = \min(f_{\text{noise}}, f_{\text{Crowther}})$ at four projection counts. Surfaces are drawn semi-transparent so the nesting is visible; lower $N_\theta$ (darker) sits inside higher $N_\theta$ (lighter). The floor lifts as $N_\theta^{1/3}$. The outer rim climbs linearly with $N_\theta$. (b)~Radial profiles of $f_{\text{eff}}(r)$ at the same projection counts (linear axes). The noise-bound central plateau and the Crowther-bound $1/r$ tail are visible. The crossover migrates outward with $N_\theta$.}
    \label{fig:sweep_ntheta}
\end{figure}

Unlike exposure, projection count $N_\theta$ moves both components of the resolution limit simultaneously. The photon-noise bound improves as $N_\theta^{1/3}$ (Equation~\ref{eq:dmin}), and the Crowther bound improves linearly, $f_{\text{Crowther}} \propto N_\theta$ (Equation~\ref{eq:crowther}). Figure~\ref{fig:sweep_ntheta} renders both effects in a single picture. Panel~(a) stacks the combined-limit 3D surface $f_{\text{eff}}(x, y) = \min(f_{\text{noise}}, f_{\text{Crowther}})$ for four projection counts $N_\theta \in \{55, 110, 220, 440\}$, drawn semi-transparent so the nesting is visible. Lower $N_\theta$ (darker) sits inside higher $N_\theta$ (lighter). The floor lifts as $N_\theta^{1/3}$ while the Crowther-dominated outer collar lifts linearly in $N_\theta$, so the outer rim climbs faster than the central plateau as projection count grows.

Figure~\ref{fig:sweep_ntheta}b plots the radial profile of $f_{\text{eff}}$ for each $N_\theta$. The profiles share the same shape, a noise-bound central plateau and a Crowther-bound $1/r$ tail at large $r$. The location and height of every feature scales with $N_\theta$, as summarised in Table~\ref{tab:ntheta_scaling}. At the lowest $N_\theta = 55$, the crossover from noise- to Crowther-bound sits at $r \approx \SI{4.5}{mm}$, well inside the \SI{13}{mm} specimen, so most of the FOV is angular-limited. As $N_\theta$ grows, the crossover migrates outward, reaching $r \approx \SI{7.1}{mm}$ at $N_\theta = 110$, $\approx \SI{11.0}{mm}$ at $N_\theta = 220$ (nominal, still inside the specimen), and at $N_\theta = 440$ the crossover has passed the rim. The whole specimen interior is now photon-noise-bound. Below this transition, $f_{\text{eff}}(r=R)$ scales linearly with $N_\theta$ (pure Crowther). Above it, scaling reverts to the $N_\theta^{1/3}$ photon-noise law.

\begin{table}[H]
\centering
\caption{Combined-limit summary across the projection-count sweep at $\Delta C = \SI{500}{HU}$, $R = \SI{13}{mm}$. ``Crossover $r$'' is the radius at which $f_{\text{noise}} = f_{\text{Crowther}}$. The dash for $N_\theta = 440$ indicates the crossover lies outside the specimen, so the entire interior is noise-bound.}
\label{tab:ntheta_scaling}
\small
\begin{tabular}{ccccc}
\toprule
$N_\theta$ & $f_{\text{noise}}(r=0)$ (lp/mm) & Crossover $r$ (mm) & $f_{\text{eff}}(r=R)$ (lp/mm) & Binding at rim \\
\midrule
55  & 1.80 & 4.5  & 0.63 & Crowther \\
110 & 2.27 & 7.1  & 1.26 & Crowther \\
220 & 2.86 & 11.0 & 2.53 & Crowther (nominal) \\
440 & 3.60 & ---  & 3.86 & photon noise \\
\bottomrule
\end{tabular}
\end{table}

The two-effect scaling makes $N_\theta$ a more efficient parameter than $N_0$ for raising low-contrast resolution, especially while the FOV is still Crowther-bound. Doubling $N_\theta$ improves the central resolution by $\sqrt[3]{2} \approx 1.26\times$ and doubles the resolution at every Crowther-bound radius. Doubling $N_0$ only delivers the first of these gains. Once the entire FOV crosses into the noise-bound regime (which happens for our \SI{13}{mm} specimen at $N_\theta \approx 288$ at this exposure), the two parameters become equivalent at every radius.

\subsection{Optimal photon allocation under a fixed dose budget}\label{sec:res_optimal}

The exposure (Section~\ref{sec:res_sweep_n0}) and projection-count (Section~\ref{sec:res_sweep_ntheta}) sweeps each vary one parameter while holding the other fixed. The scan designer is more often interested in the dual question. Given a fixed total photon budget per detector pixel
\begin{equation}\label{eq:phi}
\Phi \;\equiv\; N_0 \cdot N_\theta,
\end{equation}
how should that budget be split between $N_0$ and $N_\theta$? The optimum follows analytically.

The figure of merit is the worst-case effective resolution inside a target field of view of radius $r^*$,
\begin{equation}\label{eq:fmin}
f_{\min}(r^*) \;=\; \min_{r \,\leq\, r^*} f_{\text{eff}}(r) \;=\; \min_{r \,\leq\, r^*} \min[\,f_{\text{noise}}(r),\,f_{\text{Crowther}}(r)\,].
\end{equation}
Equation~\ref{eq:fmin} collapses to a two-term expression because $f_{\text{noise}}(r)$ is smallest at $r = 0$ (every projection ray traverses the full diameter there, and photon attenuation is maximal), and $f_{\text{Crowther}}(r)$ is smallest at $r = r^*$ (the Crowther limit falls as $1/r$). Therefore
\begin{equation}\label{eq:fmin_two_term}
f_{\min}(r^*) \;=\; \min\!\left[\,f_{\text{noise}}(0),\; f_{\text{Crowther}}(r^*)\,\right].
\end{equation}

\paragraph{The dose-resolution law.}
At the rotation axis, every projection sees the same chord $L = 2R$, so the geometric factor in Equation~\ref{eq:dmin_offaxis} collapses to a constant and Equation~\ref{eq:dmin} reduces to
\begin{equation}\label{eq:dmin_centre}
d_{\min}(0) \;=\; \left(\frac{3\,d'^2_{\text{th}}\,\Delta a_{obj}}{2\,\Phi\,e^{-2\mu R}\,\Delta\mu^2}\right)^{\!1/3}.
\end{equation}
The product $N_0 N_\theta$ enters as the single combination $\Phi$. Inverting to the spatial-frequency bound and defining the constant
\begin{equation}\label{eq:K}
K \;\equiv\; \frac{1}{2}\left(\frac{2\,e^{-2\mu R}\,\Delta\mu^2}{3\,d'^2_{\text{th}}\,\Delta a_{obj}}\right)^{\!1/3},
\end{equation}
gives
\begin{equation}\label{eq:fnoise_phi}
f_{\text{noise}}(0) \;=\; K \cdot \Phi^{1/3}.
\end{equation}
The photon-noise floor at the centre of the specimen depends only on the total photon budget, not on how it is split between $N_0$ and $N_\theta$. For our scanner at $\Delta C = \SI{500}{HU}$ inside an $R = \SI{13}{mm}$ specimen, $K = \num{0.02932}~$lp/mm$\cdot$photons$^{-1/3}$.

\paragraph{The optimum.}
With $f_{\text{noise}}(0)$ fixed by $\Phi$, the only handle on $f_{\min}(r^*)$ is $f_{\text{Crowther}}(r^*) = N_\theta / (2\Delta\theta\,r^*)$, which grows linearly with $N_\theta$. Below threshold, Crowther binds and $f_{\min}$ grows linearly in $N_\theta$. Above threshold, photon noise binds and $f_{\min}$ plateaus. The threshold is set by equating the two:
\begin{equation}\label{eq:Ntheta_star}
N_\theta^\star(r^*, \Phi) \;=\; 2\,K\,\Delta\theta\,r^* \cdot \Phi^{1/3}.
\end{equation}
The maximum achievable worst-case resolution is then
\begin{equation}\label{eq:fmin_max}
\boxed{\,f_{\min}^{\max}(r^*, \Phi) \;=\; K \cdot \Phi^{1/3},\,}
\end{equation}
attained for any $N_\theta \geq N_\theta^\star$. The corresponding per-pixel exposure is $N_0^\star = \Phi / N_\theta^\star = \Phi^{2/3} / (2 K \Delta\theta r^*)$.

\begin{figure}[H]
    \begin{center}
    \includegraphics[width=\textwidth]{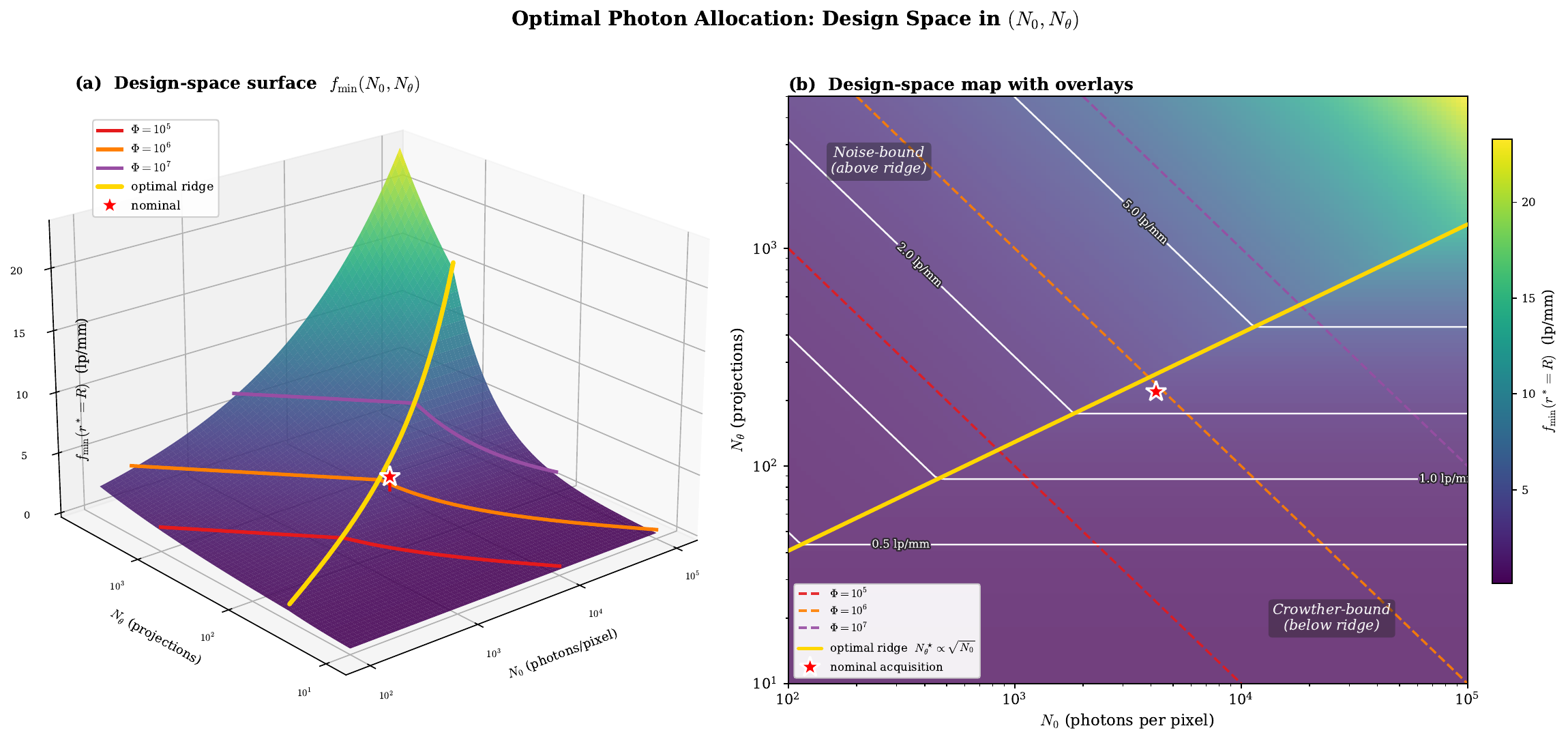}
    \end{center}
    \caption{Dose-allocation design space at $\Delta C = \SI{500}{HU}$, $r^* = R = \SI{13}{mm}$. (a)~3D surface $f_{\min}(N_0, N_\theta)$ with both axes log-spaced. Three iso-$\Phi$ trajectories (red, orange, purple for $\Phi = 10^5, 10^6, 10^7$) climb the Crowther ramp and plateau in the noise-bound regime. The red star is our nominal acquisition. (b)~2D heatmap of the same surface with design overlays: iso-$\Phi$ lines (dashed, slope $-1$), the optimal-allocation ridge $N_\theta^\star \propto \sqrt{N_0}$ (gold solid, slope $+1/2$), iso-$f_{\min}$ contours (white line), and the two regimes labelled. The nominal acquisition sits just below the ridge in the Crowther-bound half.}
    \label{fig:optimal}
\end{figure}

\paragraph{Visualisation.}
Figure~\ref{fig:optimal} renders the full design space at $r^* = R$. Figure~\ref{fig:optimal}a is the 3D surface $f_{\min}(N_0, N_\theta)$ with the two axes log-spaced. Three iso-$\Phi$ trajectories (constant total dose, in red/orange/purple for $\Phi \in \{10^5, 10^6, 10^7\}$) are drawn directly on the surface: each climbs the Crowther ramp at low $N_\theta$ and plateaus once it crosses the ridge into the noise-bound region. The plateau heights are spaced by exactly $10^{1/3} \approx 2.15$ in resolution, the dose-resolution law made visible. Figure~\ref{fig:optimal}b is the top-down 2D heatmap of the same data with the design overlays added:
\begin{itemize}
    \item \textbf{Iso-$\Phi$ lines} (dashed, slope $-1$ in log-log): trajectories of constant total dose. The scan designer's degree of freedom is to slide along one of these lines.
    \item \textbf{Optimal-allocation ridge} (gold, slope $+1/2$ in log-log): $N_\theta^\star \propto \sqrt{N_0}$ for fixed $r^*$, derived from Equation~\ref{eq:Ntheta_star} by eliminating $\Phi$. Below the ridge, Crowther binds. Above the ridge, photon noise is saturated. Walking along any iso-$\Phi$ line until it meets the ridge gives the optimal split for that dose.
    \item \textbf{Iso-$f_{\min}$ contours} (white line): constant-resolution surfaces. They are vertical (Crowther: $f_{\min}$ depends only on $N_\theta$) below the ridge and curve toward the diagonal (noise: $f_{\min}$ depends on $N_0 N_\theta$) above it.
    \item \textbf{Nominal acquisition} (red star): $(N_0, N_\theta) = (\num{4203}, 220)$, sitting visibly below the ridge, confirming that our actual scan is Crowther-limited at the rim of the specimen.
\end{itemize}

\paragraph{Practical rules.}
Three rules follow directly from Equations~\ref{eq:fnoise_phi}--\ref{eq:fmin_max}:
\begin{enumerate}
    \item \textbf{The dose-resolution law is cubic.} $f_{\min}^{\max} \propto \Phi^{1/3}$, so each doubling of worst-case resolution requires $8\times$ the total photons. This is the classical resolution-dose trade in CT~\cite{brooks_dichiro_1976, ford_fundamental_2003}, here recovered as a property of the ideal observer rather than of any specific reconstruction.
    \item \textbf{$N_\theta^\star$ scales linearly with FOV radius.} Larger ROIs demand more views in linear proportion. Imaging a 4$\times$ larger radius at the same maximum resolution requires 4$\times$ the views.
    \item \textbf{Add views when $N_\theta < N_\theta^\star$. Add flux per view when $N_\theta > N_\theta^\star$.} Adding $N_0$ when $N_\theta < N_\theta^\star$ does not improve the worst-case resolution because Crowther is binding. Symmetrically, adding $N_\theta$ when $N_\theta > N_\theta^\star$ does not help because photon noise is saturated. The optimal scan operates exactly at $N_\theta^\star$.
\end{enumerate}
\paragraph{Application to our acquisition.}
As a worked example, take our nominal acquisition with $\Phi = \num{924660}$ and $r^* = R = \SI{13}{mm}$. The optimum is $N_\theta^\star \approx 250$ views and $N_0^\star \approx 3700$ photons/pixel. Our actual acquisition uses $N_\theta = 220$ and $N_0 = 4203$, putting us slightly on the Crowther-bound side of the line but very close to it. Redistributing the same total dose toward $N_\theta^\star$ would improve the rim resolution from $f_{\text{Crowther}}(R) = \SI{2.52}{lp/mm}$ to $f_{\min}^{\max} = \SI{2.86}{lp/mm}$, a $\sim 14\%$ gain rather than the factor-of-two-plus gap that would exist at larger specimen radii. The constraint we omit is the detector noise floor. At very low $N_0$ per pixel, electronic noise stops the Poisson assumption from holding, and Equation~\ref{eq:dmin_centre} ceases to apply.

\subsection{Interactive web calculator}\label{sec:res_webcalc}

The closed-form bounds derived above are implemented in an open-source browser-based calculator~\cite{wiegmann_ideal_observer_calculator_2026} released with this work. It accepts arbitrary scanner geometries (any $N_0$, $N_\theta$, arc coverage, source-to-isocentre distance, contrast, and specimen geometry) and renders the photon-noise (Equations~\ref{eq:dmin}--\ref{eq:dmin_offaxis}), Crowther, and combined-limit maps. Live deployment: \url{https://ubc-ford-lab.github.io/ideal_observer_SKE_BKE_model/}.

\section{Discussion}\label{sec:discussion}

\subsection{Scope of each limit and learned-prior reconstructions}\label{sec:disc_scope}

Section~\ref{sec:res_bridge} laid out the formal scope of the two bounds. Photon noise is universal (a property of the projection data themselves), while Crowther is specific to reconstructions without strong signal priors. Two implications follow. First, the gap between actual reconstructions and the ideal-observer bound (Section~\ref{sec:res_ideal}) is the room available for improvement on a fixed dataset. Closing it requires algorithms that match the matched filter on the noise side, not more dose. Second, the Crowther line in Figure~\ref{fig:combined} should not be read as a strict ceiling for every reconstruction class. Learned reconstructions that incorporate signal priors~\cite{zang_intratomo_2021, ruckert_neat_2022, shen_nerp_2022, ulyanov_dip_2018, mildenhall_nerf_2020} have been demonstrated to recover frequencies above $f_{\text{Crowther}}$ in sparse-view CT. The cost is the assumption that the prior is correct for the signal at hand. Where it fits, the assumption is benign. For out-of-distribution anatomy or novel pathology, there is no such guarantee. The photon-noise component alone is the binding floor that applies regardless of prior strength. More broadly, because the binding limit varies spatially across the field of view, the absolute resolution numbers reported here should be read as approximations. The framework is most useful for relative comparison between reconstructions on a fixed dataset, where the algorithm is the only variable.

\subsection{Scan-design implications}\label{sec:disc_design}

The parametric sweeps in Sections~\ref{sec:res_sweep_n0}--\ref{sec:res_sweep_ntheta} translate into the following design rules. For an experimenter constrained by total dose $D_{\text{tot}} \propto N_0 \cdot N_\theta$:
\begin{itemize}
    \item If photon noise dominates over the region of interest (low contrast, central placement of the specimen), both $N_0$ and $N_\theta$ buy resolution as $\propto D_{\text{tot}}^{1/3}$. The choice between more views and brighter views is then governed by mechanical / electronic constraints (gantry stability, detector saturation, electronic noise floor), not by the resolution bound itself.
    \item If angular sampling dominates (high contrast, peripheral specimen placement), $N_\theta$ is strictly preferred to $N_0$. It improves both bounds, while $N_0$ improves only the noise bound that is not binding.
\end{itemize}
For our scanner with low-contrast targets, the binding limit is photon noise everywhere in the typical specimen volume (Figure~\ref{fig:combined}b), so the practical takeaway is that low-contrast resolution is dose-limited, not geometry-limited, on this system at the nominal acquisition. Centring the specimen on the rotation axis is still recommended, because it keeps the angular limit safely above the noise limit, but the gains are second-order at low contrast.

\begin{figure}[H]
    \begin{center}
    \includegraphics[width=0.95\textwidth]{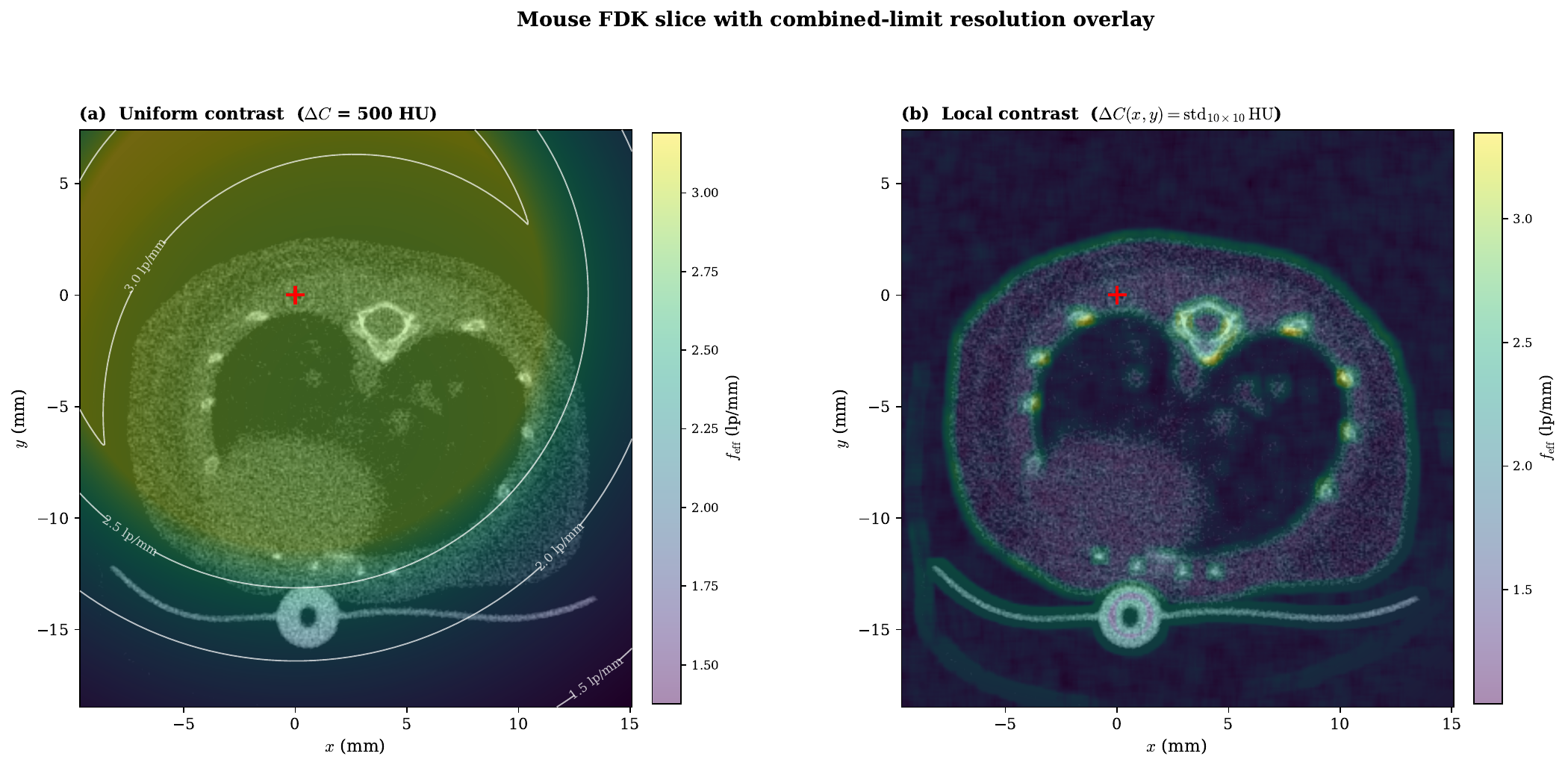}
    \end{center}
    \caption{Pixel-by-pixel application of the combined-limit map to an FDK reconstruction of a representative mouse specimen. (a)~Uniform-contrast assumption $\Delta C = \SI{500}{HU}$ everywhere. $f_{\text{eff}}$ varies only by a factor of $\sim 2$ across the FOV ($\sim$\SIrange{1.4}{3.2}{lp/mm}). (b)~Anatomy-aware contrast: $\Delta C(x, y)$ is the local HU standard deviation over a $10\times 10$~pixel window, i.e.\ the contrast scale of features actually present at each pixel. The map now answers ``can the features that live here be resolved?'' rather than the under-specified ``what is the resolution at this pixel?''. High-contrast structures (bone, lung--air, tubing) glow at high $f_{\text{eff}}$. Homogeneous tissue interiors sit lower. Each panel uses its own linear colorbar. Specimen model $R = \SI{13}{mm}$, $N_\theta = 220$ over \ang{192}.}
    \label{fig:mouse_overlay}
\end{figure}

Because $f_{\text{eff}}(x, y)$ is defined everywhere in the field of view, the bound can be sampled at the actual pixel locations of a reconstructed specimen to give a resolution map over real anatomy. Figure~\ref{fig:mouse_overlay} illustrates this on an FDK reconstruction of a representative mouse, with the combined limit overlaid transparently in two variants.

\paragraph{Caveat: the example is illustrative, not quantitative.}
The figure mixes two physical models that are not in fact consistent with each other. The photon-noise component $f_{\text{noise}}$ is evaluated using the uniform cylindrical water-equivalent specimen model of Section~\ref{sec:res_3d} ($R = \SI{13}{mm}$ of homogeneous tissue), but the underlying anatomy is a real mouse thorax with air-filled lungs, bone, and soft tissue with sharply different linear attenuation. The line integrals, and so the actual transmitted photon counts $N_{bg}$ reaching each pixel, differ from those of the uniform cylinder. Sometimes substantially. Air-filled lungs let through far more photons than the cylinder model assumes, and vertebrae considerably fewer. A quantitatively correct map would forward-project the anatomical $\mu$-map at every projection angle to recover the true per-pixel $N_{bg}$, then plug into Equation~\ref{eq:dmin_offaxis}. We have not done that here. Figure~\ref{fig:mouse_overlay} illustrates how the bound can be combined with reconstructed anatomy. It is not a quantitatively accurate detectability map for this specific specimen.

With that scope in mind, Figure~\ref{fig:mouse_overlay}a evaluates $f_{\text{eff}}$ at a fixed contrast $\Delta C = \SI{500}{HU}$ everywhere, isolating the geometric and noise structure of the bound itself. Resolution varies only by a factor of $\sim 2$ across the FOV, with the variation set by chord length and radial distance from the rotation axis.

Figure~\ref{fig:mouse_overlay}b goes further. At each pixel, the contrast input to the formula is set to the local intensity standard deviation $\Delta C(x, y) = \mathrm{std}_{w}[\mathrm{HU}](x, y)$ measured in a small $w = 10 \times 10$~pixel ($\approx \SI{0.75}{mm}$) window. The interpretation is anatomically grounded: $\Delta C(x, y)$ is the contrast scale of features actually present in the local neighbourhood, so the map answers ``can the features that live at this pixel be resolved here?'' rather than the under-specified ``what is the resolution at this pixel?''. Resolution is contrast-dependent in CT. No single number exists without a contrast specification. The map highlights edges and high-frequency anatomical structures. Bone surfaces, lung-air interfaces, and the small tubing at the bottom of the FOV glow at high $f_{\text{eff}}$, while homogeneous soft-tissue interiors sit lower because the local contrast scale there (dominated by reconstruction noise, $\sim$\SI{75}{HU}) is much smaller than at edges. A pixel-by-pixel map answers questions that a single MTF number cannot, such as which anatomical region to use for quantitative endpoints, where to place an ROI for a measurement that needs a specific feature size, and whether a lesion of known contrast and location is detectable in the data at hand.

\subsection{Limitations and unmodelled sources of resolution loss}\label{sec:disc_limitations}

The closed-form bound keeps only two limits and ignores several physical effects that contribute to real-world resolution loss. For each, we state the direction in which it would shift the bound.

\paragraph{SKE/BKE optimism.}
The ideal observer assumes the signal location and background are known exactly. Real detection tasks face structured anatomy, ``anatomical noise'', that our model does not include. The generalised detectability analysis of Gang and colleagues~\cite{gang_anatomical_2010}, building on Hanson's foundational detectability work~\cite{hanson_detectability_1979}, shows that anatomical clutter can degrade detectability substantially in cone-beam CT, with the magnitude depending on the task, the clutter spectrum, and the acquisition geometry. The bound here is an upper limit on detectability. Real performance, even for an ideal observer, will be worse than reported.

\paragraph{Electronic noise floor and polyenergetic spectrum.}
The Poisson assumption breaks at very low per-pixel photon counts where electronic noise dominates. Section~\ref{sec:res_optimal} mentions this as the floor on the $N_0$ side of any dose redistribution. The polyenergetic X-ray spectrum and beam hardening additionally make $\mu$ depend on path length. For our setup this affects $N_{bg}$ by a few percent and does not change the scaling laws.

\paragraph{Cone-beam Tuy-condition error.}
The 2D fan-beam off-axis formula (Equation~\ref{eq:dmin_offaxis}) neglects the FDK approximation error that grows with cone angle. For our \ang{4} cone half-angle Section~\ref{sec:res_3d} shows this is sub-percent. Wide-detector clinical CBCT geometries with \ang{10}--\ang{15} cone would require an explicit correction.

\paragraph{Relation to the cascaded-systems / NEQ framework.}
A more rigorous treatment of all the above effects together is the Fourier-domain cascaded-systems analysis of Rabbani, Shaw \& Van Metter~\cite{rabbani_cascade_1987}, Cunningham, Westmore \& Fenster~\cite{cunningham_cascaded_1994}, and Tward \& Siewerdsen~\cite{tward_cascaded_2008}, which derives a generalised noise-equivalent quanta NEQ($f$) accounting for photon statistics, detector blur (MTF $T(f)$), focal-spot blur, additive electronic noise (NPS $S_e(f)$), and reconstruction filtering. The closed-form $d_{\min}$ used here is the projection-domain matched-filter limit of that machinery, evaluated before detector blur, electronic noise, and reconstruction filtering enter the chain. In this limit, NEQ($f$) reduces to the input photon count $N_0$, and the matched-filter detectability recovers our $d^3$ disc-detection scaling, consistent with the cascaded results of Gang et al.~\cite{gang_anatomical_2010}. Detector MTF rolloff at the spatial frequencies the task depends on, electronic-noise dominance at very low $N_0$ per pixel, and reconstruction-filter shaping are captured by full cascaded-NEQ evaluation and should be used when these effects are non-negligible. The closed forms suit scan-design trade-off questions, and a further advantage over a full cascaded-NEQ evaluation is that their spatially varying structure can be evaluated and displayed directly, as in the combined-limit maps above. Absolute predictions in regimes where any of these effects matter should be cross-checked against the cascaded NEQ value.

\section{Conclusion}\label{sec:conclusion}

Spatial resolution in preclinical cone-beam micro-CT is bounded by two reconstruction-independent limits. A photon-noise limit applies to any reconstruction algorithm. An angular-sampling limit applies to reconstructions without signal priors. The two combine into a spatially varying resolution surface that depends only on scan parameters and feature contrast, not on the choice of reconstruction. For our GE eXplore CT~120 on a typical mouse-specimen geometry ($R = \SI{13}{mm}$), the photon-noise limit at the rotation axis is \SI{0.98}{lp/mm} at $\Delta C = \SI{100}{HU}$ and \SI{2.86}{lp/mm} at \SI{500}{HU}. Classical FDK and iterative reconstructions operate several times above this bound, leaving headroom for noise-side algorithmic improvement. Reconstructions that incorporate signal priors may exceed the Crowther limit at the cost of relying on those priors. They cannot exceed the photon-noise limit at any cost.

\bibliographystyle{unsrtnat}
\bibliography{preprint}

\newpage

\section*{Acknowledgments}
This work was supported by the BC Lung Foundation and by a Natural Sciences and Engineering Research Council of Canada (NSERC) Discovery Grant (RGPIN-2024-06330) to N.L. Ford.

\section*{Author Contributions}
Falk L. Wiegmann and Nancy L. Ford contributed to the research direction and conceptualisation. Falk L. Wiegmann derived the closed-form bounds, implemented the numerical evaluation and the interactive web calculator, performed the analysis, and wrote the manuscript. Nancy L. Ford supervised the research, secured funding, provided critical review, and edited the manuscript.

\section*{Competing Interests}
The authors declare no competing interests.

\section*{Data Availability}
The interactive web calculator and its source code are publicly available at \url{https://github.com/UBC-Ford-lab/ideal_observer_SKE_BKE_model} with a live deployment at \url{https://ubc-ford-lab.github.io/ideal_observer_SKE_BKE_model/}. The figure-generation scripts are available from the corresponding author on reasonable request. Scan data are available from the corresponding author on reasonable request.

\section*{Ethics Statement}
No animal experiments were conducted as part of this study. All scan data used here were collected as part of previously published studies~\cite{ford_respiratory_2025, ford_spie_2023} under protocols approved by the University of British Columbia Animal Care Committee (Protocol No.\ A21-0060, approved August 31, 2021) and performed in accordance with the ARRIVE guidelines~\cite{percie_du_sert_arrive_2020}. The imaging data were re-used here for reconstruction-bound evaluation only. No animals were imaged, handled, or subjected to any procedures as part of this work.

\section*{Use of AI Tools}
Claude (Anthropic) was used in some places to assist with manuscript preparation. All AI-generated content was reviewed, verified, and revised by the authors, who take full responsibility for the final manuscript.

\end{document}